\documentstyle[11pt]{article}
\newcommand{\be}{\begin{equation}}
\newcommand{\ee}{\end{equation}}
\newcommand{\ba}{\begin{eqnarray}}
\newcommand{\ea}{\end{eqnarray}}
\newcommand{\bann}{\begin{eqnarray*}}
\newcommand{\eann}{\end{eqnarray*}}

\begin{document}
\hbadness=10000
\setcounter{page}{1}
\title{
\vspace{-3.0cm}
\hspace{-2.0cm}
{\huge \bf 
$\pi^-/\pi^+$ ratio in heavy ions collisions: Coulomb effect or
chemical equilibration ?}
}
\author{
N. Arbex$^1$\thanks{E. Mail: arbex@mailer.uni-marburg.de},
U. Ornik$^2$\thanks{E. Mail: ornik@warp.soultek.de},
M. Pl\"umer$^1$\thanks{E. Mail: pluemer@mailer.uni-marburg.de},
B.R. Schlei$^3$\thanks{E. Mail: schlei@t2.lanl.gov} and
R.M. Weiner$^1$$^4$\thanks{E. Mail: weiner@mailer.uni-marburg.de}
}
\date{$^1$ Physics Department, Univ. of Marburg, Germany \\
      $^2$ Soultek Internet Service, Marburg, Germany\\
      $^3$ Theoretical Division, Los Alamos National Laboratory, USA\\
      $^4$ Laboratoire de Physique Theorique et Hautes \'Energies, Univ. 
Paris-Sud, France}
\maketitle
\begin{abstract}
We calculate the $\pi^-/\pi^+$ ratio for $Pb+Pb$
at CERN/SPS energies
and for $Au+Au$ at BNL/AGS energies
using a (3+1) dimensional hydrodynamical model.  
Without consideration of Coulomb effect
an enhancement of this ratio at low $m_t$
is found compatible with that observed in these experiments.
Our calculations are based on previous (3+1) dimensional hydrodynamical
simulations (HYLANDER), which described many other aspects of experimental
data.
In this model the observed enhancement 
is a consequence of baryon and 
strangeness conservation and of chemical equilibration of the 
system and is caused by the decay of produced hyperons,
which leads to a difference in the total number
of positive and negative pions as well.
Based on the same approach, we also present results 
for the $\pi^-/\pi^+$ ratio
for $S+S$ $(CERN/SPS)$ collisions, 
where we find a similar effect. 
The absence of the enhancement of the $\pi^-/\pi^+$ ratio
in the $S+S$ data presented
by the NA44 Collaboration, if confirmed, 
could indicate that chemical 
equilibration has not yet been estabilished in this reaction. 

\end{abstract}
 
\newpage

Recently the NA44 Collaboration has
presented results
of measurements of $\pi^-/\pi^+$ ratios in heavy
ions reactions at the CERN/SPS accelerator at 
incident beam energies of 158
and 200 GeV/A \cite{coulomb}.

The observed excess of negative over positive pions
in the low $m_t$ region
was interpreted in this report as due to Coulomb 
final state interactions, although no quantitative estimate
of this effect has been given. Important arguments in this 
interpretation were the following:
\\
(i) RQMD predictions, including the decays of resonances,
could not account for this excess.
\\
(ii) in $Pb+Pb$ reactions the effect is 
more pronounced than in $S+S$ reactions.

In this letter we present a calculation of the pion ratios
and of related particle yields using a (3+1) dimensional
hydrodynamical model (HYLANDER) \cite{ornik} for heavy ion collisions 
at CERN/SPS and BNL/AGS energies.
In previous publications on
200 $AGeV$ $S+S$, 158 $AGeV$ $Pb+Pb$
and 11 $AGeV$ $Au+Au$ we described many different physical observables
concerning these reactions
such as single inclusive spectra and pion correlations
(see \cite{Bo}-\cite{ags-all}).
The present calculation is based on these previous simulations,
where also resonance decays were taken into account.
Since we consider central collisions, we assume axial symmetry
around the beam direction.

For the initial conditions of the fireball at SPS energies we
had to take into account a certain degree of transparency of the 
colliding nuclei,
whereas for the reaction at $AGS$ energies we considered 3-d full-stopping
from the moment of impact (for more details see \cite{Bo} and
\cite{ags-all}).

All the results presented 
here have been obtained by using
an equation of state based on lattice QCD calculations,
exhibiting a phase transition from a quark-gluon
plasma to a hadronic phase at $T_c=200$ $MeV$ 
\cite{ornik}\cite{lat}. The freeze-out temperature is 
chosen as $T_f=139$ $MeV$.

In our calculations for $SPS$ energies we took into account 
the detector acceptance as defined in \cite{coulomb} and \cite{expcond}
\footnote{We find, e.g., that  
only about $17\%$ of negative pions from $\Lambda$ decay and 
about $15\%$ of negative pions from $\Sigma$ decay 
survive the detection conditions
for the $Pb+Pb$ reaction. For the $S+S$ reaction the numbers are
respectively $23\%$ and $22\%$. However,
the presented results for the pion ratio
are not strongly affected by the limited detector acceptance.}.

In Fig. 1 we show our results for $Pb+Pb$ and $S+S$
collisions at SPS energies. They are compared to the data published 
in reference \cite{coulomb}. 
For $Pb+Pb$ collisions the
simulation
is compatible with the data (Fig.1(a)),
while for $S+S$ collisions (Fig.1(b)) this is not the case:
here our model predicts
an enhancement as well,
which is apparently not 
present in the data. 

In Fig. 2 we show the $\pi^-/\pi^+$ ratio for $Au+Au$ collisions
at $AGS$ energies \footnote{For more details concerning this 
simulation cf. \cite{ags-all}}.
The results of our simulation are compared with
preliminary data from reference \cite{QM95}.
Here the ratio reaches even bigger values than those found 
for $Pb+Pb(SPS)$ data. Below we will discuss  
possible reasons for this.

In our model, the low $m_t$ enhancement in the $\pi^-$ production
is a consequence of nuclear stopping, thermalization, 
hadronization and chemical equilibration of the fireball
produced in a relativistic heavy-ions collision.
At the beginning, a large number of baryon stopped
in the central region will thermalize.
This induces a strange chemical potential
which favours the production of hyperons (which are not
present in the initial state). The number of hyperons reaches 
a maximum value if the equilibration is complete.
After hadronization (freeze-out) the hyperons decay dominantly into
$\pi^- + (p,n)$ channels. They are concentrated in the soft 
$m_t$ region because of the low amount of available
kinetic energy in the hyperon decay.

We already analysed the enhancement in the $\pi^-$ production for 
$Au+Au$ in \cite{ags-all}. 
We showed that
taking into account baryon and strangeness conservation
as well as strangeness equilibration,
including the decay of resonances 
in the final stage, 
the difference $N_{\pi^-} - N_{\pi^+}$ 
in our model is determined by the amount of produced hyperons,
i.e.,
\be
N_{\pi^-}-N_{\pi^+}= N_{\mbox{ \footnotesize hyperons 
decaying into } \pi^-}-
N_{\mbox{ \footnotesize hyperons decaying into } \pi^+}
\ee
 We have \cite{ags-all}:
\be
N_{\pi^-} - N_{\pi^+} \simeq 
0.64 N_{\Lambda} 
+ N_{\Sigma^-(1190)} + 0.64 N_{\Sigma^0} - 0.48 N_{\Sigma^+}
+ 1.64 N_{\Xi^-} + 0.64 N_{\Xi^0}
\mbox{             }\footnote{In this simulation 
we are considering resonances 
with masses up to $1.5$ $GeV$.
About $80 \%$ of the enhancement of the $\pi^-/\pi^+$ ratio 
is due to lambdas.}
\ee
On the other hand, if the Coulomb effect would be the main
mechanism in the observed excess and the resonance contribution
would be of secondary importance, then
the total number of $\pi^-$ should be almost equal to the
total number of $\pi^+$ and consequently there should appear at large
$m_t$ a compensating $\pi^+/\pi^-$ excess.
This means that the number
of pions ( or the pion ratio) at large $m_t$ can help to distinguish between
these two interpretations.
Interestingly enough, up to $m_t-m_{\pi}=0.8$ $GeV$ the available
$Au+Au$ data do not show an excess of $\pi^+$ over $\pi^-$.
They also show that the total number
of $\pi^-$ is significantly larger than that of $\pi^+$.
($N_{\pi^-}-N_{\pi^+}\approx40$ \cite{QM95})
(For $Pb+Pb$ no such data are yet avaiable).

We also find that the enhancement of the pion ratio
depends on the final baryon density (and therefore
on the final baryon chemical potential) of the fireball
at freeze-out.
The computed baryon density values at freeze out are: 
0.072 $n_0$, 0.094 $n_0$ and 0.192 $n_0$
for $S+S(SPS)$, $Pb+Pb(SPS)$ and $Au+Au(AGS)$ 
respectively
\footnote{$n_0$ is the normal baryon density $=0.14/fm^3$.}.
The reduction of the enhancement from AGS to
SPS energies can therefore be interpreted as due to an increase of the
transparency effect.

One should stress that the $K^-/K^+$
and $\bar{p}/p$ ratios as predicted by our model
do not depend on $m_t$, in agreement with the results
cited in \cite{coulomb}.
Furthermore the same approach predicted and/or reproduced
correctly the rapidity and transverse momentum spectra
of protons and negative hadrons
as well as the Bose-Einstein correlations in both
$S+S$ and $Pb+Pb$ reactions at SPS energies.
In $Au+Au$ at AGS energies it reproduces all
single inclusive data (protons, negative and positive
pions and kaons). 

A possible interpretation of our results
for $Pb+Pb$ reactions is that the Coulomb effect 
is in fact much smaller than expected in \cite{coulomb}
and chemical equilibration has to be taken into account
as an alternative explanation.

Finally we would like to comment on the fact
that our model overestimates
the 200 $AGeV$ $S+S$ pion ratio 
compared to the avaliable data. 

In \cite{Bo} we calculated the lambda production for
$S+S(SPS)$ as presented by the NA35 Collaboration 
in \cite{NA35-lam}.
The multiplicity calculated in our approach was $8.77$ and 
the experimental value was $8.2\pm0.9$
\footnote{The NA35 Collaboration presented 
in a later paper \cite{NA35-lam2} a new value for the $\Lambda$
total multiplicity: $9.4\pm1.0$. In this experiment the
$\Sigma^0$ were detected together with the $\Lambda$,
so as in our calculation presented in \cite{Bo}.}.

The overestimation in the pion ratio
is surprising in so far
as there appears to be agreement between the calculated
$\Lambda$ rate and the measured one (which
suggests that a similar agreement might hold
for the other, not yet measured, hyperon rates) and this
would necessarily imply an excess of produced $\pi^-$
compared with $\pi^+$ (see eq.(2)). 
However 
our model is based on the hypothesis of chemical equilibration 
at freeze-out, and 
the overestimate of the
pion ratio for $S+S$ might indicate that a complete 
chemical equilibration
in this system is not  reached until freeze-out, whereas
it is in $Pb+Pb$ and $Au+Au$.
One expects that the bigger the system or the longer
the life-time 
$( \tau_{(Pb+Pb)} \approx$ $14$ $fm/c$, $\tau_{(S+S)} \approx$ 
$7$ $fm/c$, $\tau_{(Au+Au)} \approx$ $10$ $fm/c )$ \cite{Bo}
\cite{blei1}\cite{ags-all} 
the higher the
degree of chemical equilibration at freeze-out.
This would indicate that the
$\pi^-/\pi^+$ ratio is more sensitive to the establishment
of chemical equilibrium than other physical observables.
However we believe that such a conclusion,
althought extremely interesting and of possible practical
value, might be premature, because it strongly depends on
the accuracy of preliminary $S+S$ data.

In order to obtain a clarification of the issues raised
above the following experimental steps appear necessary:
\\
(a) Determination of all accessible hyperon rates
in $S+S(SPS)$, $Pb+Pb(SPS)$ and $Au+Au(AGS)$ reactions.
\\
(b)Measurements of the $\pi^-/\pi^+$ ratio at large $m_t$ in 
all three reactions.
\\
(c) Remeasurements of the $\pi^-/\pi^+$ ratio in $S+S(SPS)$
at low $m_t$.
\\
\\
\\
We are indebted to J. Dodd, N. Xu, A. L\"orstad and J. Sullivan
for helpful discussions.
This work was supported in part by the CNPq (Brazil-Brasilia),
by the Deutsche Forschungsgemeinschaft (DFG) and University
of California.

\newpage

{\huge\bf Figure Captions}

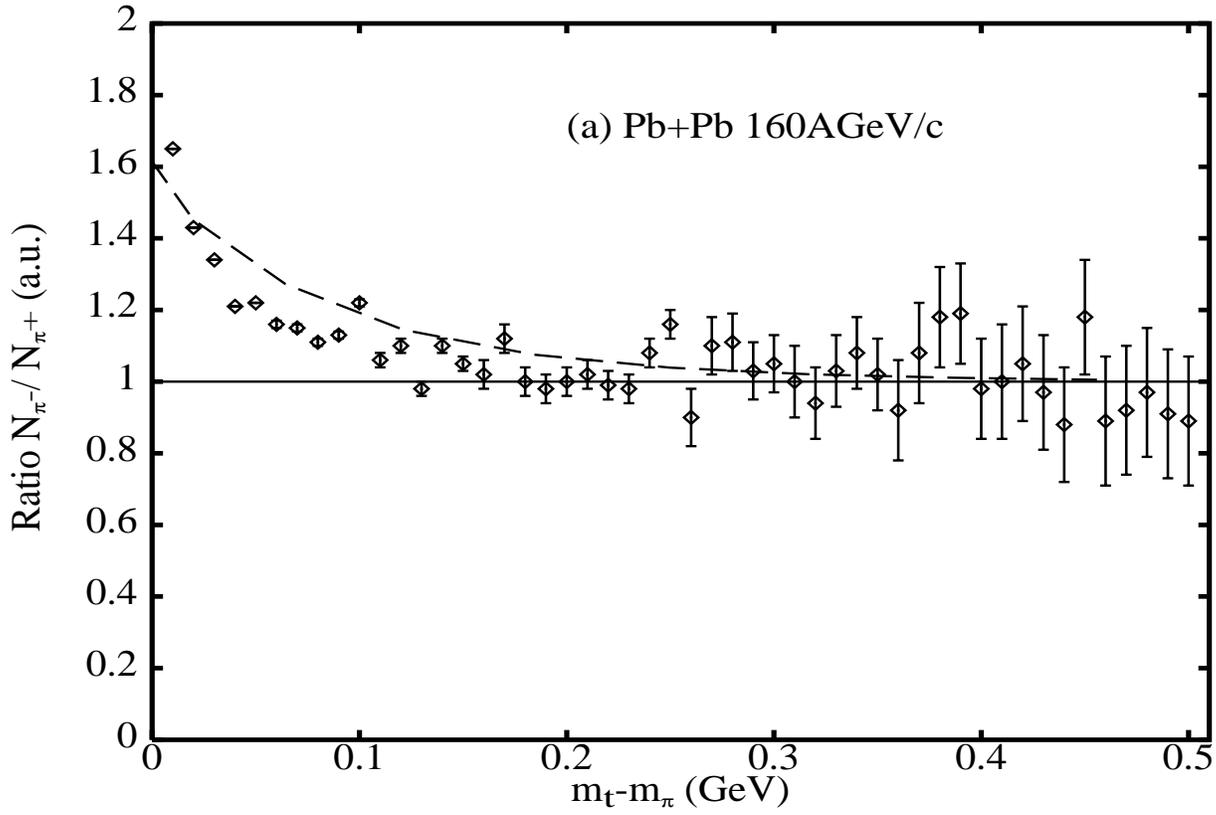
{\bf Fig. 1:} $\pi^-/\pi^+$ ratio obtained 
from a (3+1) dimensional hydrodynamical simulation (HYLANDER) in
comparison to the data from ref.\cite{coulomb},
(a) for $Pb+Pb$ collision and (b) for $S+S$ collision.
Both cases refer to CERN/SPS energies.
 
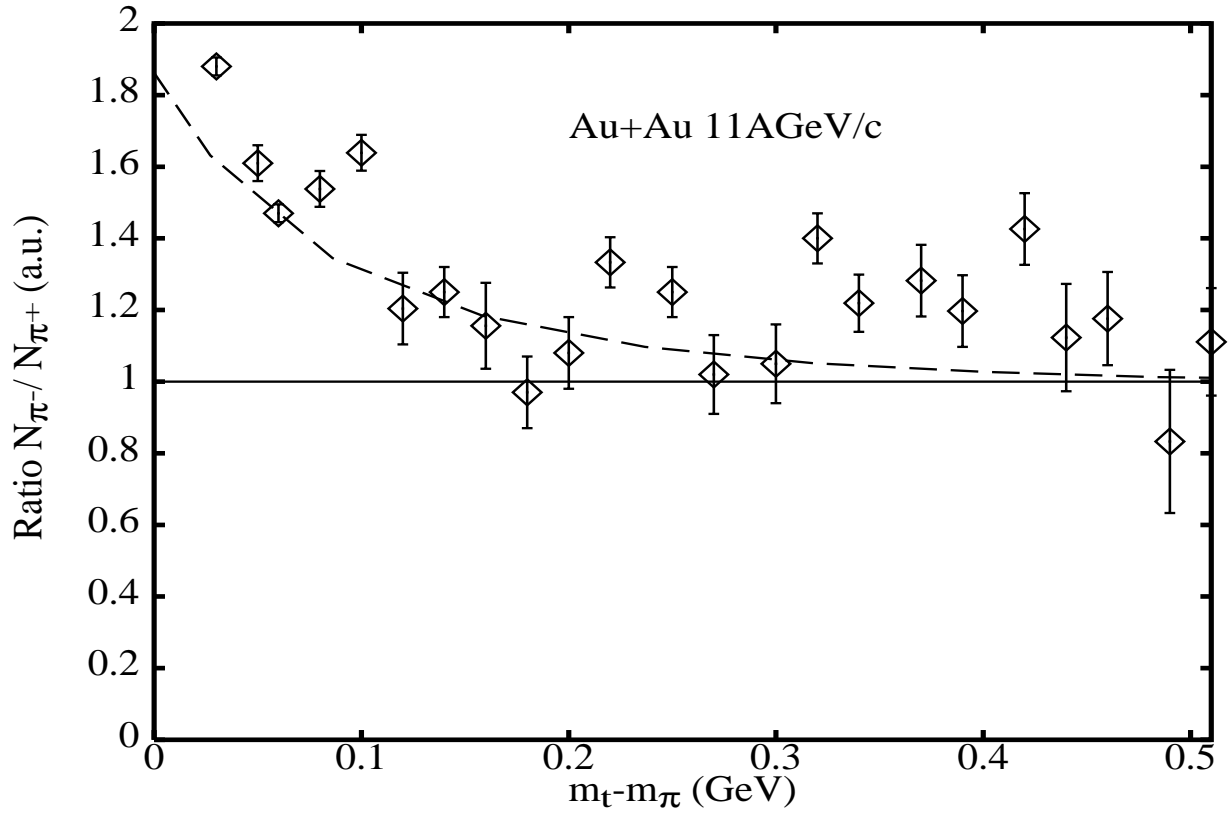
{\bf Fig. 2:} $\pi^-/\pi^+$ ratio obtained
from a (3+1) dimensional hydrodynamical simulation (HYLANDER) in
comparison to the data from ref. \cite{QM95}
for $Au+Au$ collision at BNL/AGS energies.

\newpage

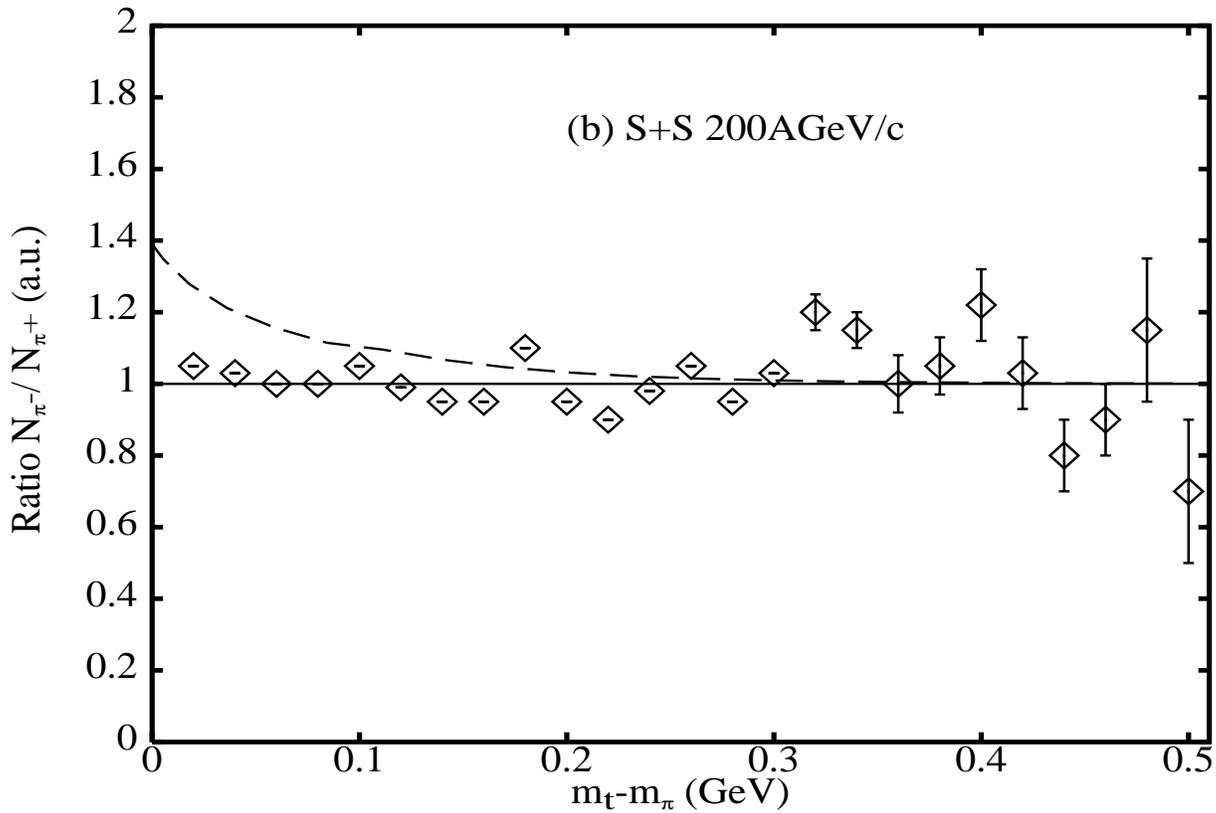
\end{document}